# Noise thermometry and electron thermometry of a sample-on-cantilever system below 1 Kelvin

A. C. Bleszynski-Jayich[1], W. E. Shanks[1] and J. G. E. Harris[1,2]

[1] Department of Physics, Yale University, New Haven CT, 06520, USA
[2] Department of Applied Physics, Yale University, New Haven CT, 06520, USA

We have used two types of thermometry to study thermal fluctuations in a microcantilever-based system below 1 K. We measured the temperature of a cantilever's macroscopic degree-of-freedom (via the Brownian motion of its lowest flexural mode) and its microscopic degrees-of-freedom (via the electron temperature of a metal sample mounted on the cantilever). We also measured both temperatures' response to a localized heat source. We find it possible to maintain thermal equilibrium between these two temperatures and a refrigerator down to at least 300 mK. These results are promising for ongoing experiments to probe quantum effects using micromechanical devices.

There are at least two distinct "temperatures" relevant to the performance of mechanical devices. The first is the effective temperature associated with the device's Brownian motion. This thermo-mechanical noise temperature, $T_n$, sets a fundamental limit to the device's force sensitivity. It is relevant in magnetic resonance force microscopy (MRFM)[1,2], atomic force microscopy (AFM), and torque magnetometry[3-5]. It also sets limits on the observation of quantum effects in mechanical oscillators[6-9]. As a result there is considerable interest in lowering this "Brownian" temperature via cryogenics[10-13] and/or cold damping techniques such as laser cooling[14-18].

The second important temperature is that of the cantilever's microscopic degrees of freedom. For sample-on-cantilever experiments, this sets the temperature of the sample attached to the cantilever, $T_e$ [3-5,19,20] and is important for MRFM and torque magnetometry experiments.

In principle, both $T_n$ and $T_e$ can be lowered by placing the cantilever in contact with a thermal bath (i.e., a refrigerator) at temperature $T_b$. However thermal equilibrium between the bath, the lever's Brownian motion, and a sample affixed to the lever is not assured. Factors preventing equilibration include the extreme aspect ratio of typical cantilevers, the insulating nature of most cantilever materials, and the injection of heat by the lever's readout mechanism (e.g., a laser).

Previous experiments have studied $T_e$ of a sample at the end of a gold-coated cantilever between 4 K and 16 K [19]. In other experiments, $T_n$ has been cooled by a fridge to 200 mK[10] in micromechanical systems and to 56 mK in nanomechanical devices[12]. We are not aware of any direct measurements of both $T_n$ and $T_e$ in a single system.

Here we present measurements of $T_n$ of a cantilever and $T_e$ of an aluminum grain attached to the end of the cantilever. $T_n$ is measured via the cantilever's Brownian motion, while $T_e$ is measured via the grain's superconducting critical field $H_c$. We also measure the response of $T_n$ and $T_e$ to the laser interferometer which monitors the cantilever. We find that $T_n$ and $T_e$ remain in good contact with each other and with $T_b$ for temperatures down to 300 mK and laser powers $P_{inc}$ below ~ 25 nW. At higher laser powers $T_e$ and $T_n$ increase above $T_b$ in a manner consistent with diffusive phonon-mediated heat transport through the cantilever.

These experiments were performed in a $^3$He refrigerator[21]. A schematic of our setup is shown in Fig. 1(a). A single crystal silicon cantilever[22] of length $L$ = 500 µm, width $w$ = 100 µm, thickness $\tau$ = 1 µm, and doping ~ $10^{18}$ cm$^{-3}$ is mounted on a piezoelectric actuator and thermally linked to the refrigerator. A fiber optic interferometer is used to measure the cantilever deflection $x$. This interferometer is formed between the cantilever and the cleaved face of a single mode optical fiber ~100 µm from the cantilever. The interferometer uses a laser wavelength $\lambda$ = 1550 nm.

The noise temperature is determined by measuring the mean square displacement $\langle x^2 \rangle$ of the cantilever's free end. From the equipartition theorem $T_n = k \langle x^2 \rangle / k_B$ where $k_B$ is the cantilever's spring constant. To obtain an absolute measurement of the displacement $x$, we calibrate the interferometer signal by applying a sinusoidal drive to the piezo actuator and measuring the fundamental Fourier component of the interferometer signal on a lock-in amplifier as a function of the drive amplitude. The data are shown in Fig. 2 (a).

To fit this data we note that the optical field at the photodiode has two sources: light reflected from the fiber's end $E_1$ (which we assume is constant) and light reflected from the cantilever $E_2$ (we take $E_1$ and $E_2$ to be complex). As the cantilever deflects the phase of $E_2$ changes, producing the interferometric signal. The cantilever deflection also modulates the amplitude of $E_2$ since the amount of reflected light coupled back into the fiber varies with the cantilever's angle relative to the fiber axis. This effect is small enough to be expanded to first order in $x$. Thus we can write the total field at the photodiode as $E_{tot} = E_1 + E_2^{(0)}(1+\varepsilon\{x(t)-x_0\})e^{2ikx(t)}$ where $k = 2\pi/\lambda$, $\varepsilon \equiv \partial E_2/\partial x |_{x=0}$, $\varepsilon x(t) \ll 1$, $x_0$ is the equilibrium position of the cantilever, and $E_2^{(0)}$ is the value of $E_2$ for $x = x_0$. The time dependent cantilever position is $x(t) = x_0 + x_1 \sin(2\pi f t)$ where $x_1$ is the amplitude of the cantilever's oscillation and $f$ is its frequency. The lock-in signal $V_{lockin}$ is proportional to the Fourier component of $|E_{tot}|^2$ at $f$:

$$V_{lockin} \propto 2E_2\varepsilon x_1 - 4E_1 E_2 \sin(2kx_0) J_1(2kx_1) + 2E_1 E_2 \varepsilon \cos(2kx_0) x_1 \{J_1(2kx_1) - J_2(2kx_1)\}$$

where we have kept terms linear in $\varepsilon$. Fitting the data in Fig. 2 (a) to this expression allows us to convert $V_{lockin}$ to an absolute displacement $x$ in terms of the known laser wavelength.

We then monitor the interferometer signal when no drive is applied to the piezo actuator and use the calibration described above to convert this signal to $S_x$, the power spectral density of the cantilever's undriven motion (Fig. 2 (b)). The data are fit to the response function of a damped harmonic oscillator, giving a quality factor $Q = 70,000$ and resonant frequency $f_0$ = 7575 Hz. The baseline in Fig. 2 (b) is a factor of 4 above the photon shot noise. We note that the cantilevers used in the $T_n$ measurements did not have an Al grain attached.

The area under the fit in Fig. 2 (b) (after subtracting the baseline) is $\langle x^2 \rangle$. We measured $\langle x^2 \rangle$ at fridge temperatures between $T_b$ = 300 mK and 4.2 K (Fig. 2 (c)). The linear dependence of $\langle x^2 \rangle$ on $T_b$ and its extrapolation to zero at $T_b$ = 0 K confirm that the force noise driving the cantilever is thermal and hence that the motion in Fig. 2 (b) is Brownian. Importantly, Fig. 2 (c) indicates that $T_n$ remains in equilibrium with the refrigerator down to 300 mK for $P_{inc}$ = 25 nW.

To determine the electron temperature of a sample on a cantilever, we measure a nominally identical cantilever on the same chip to which we attached a ~10 μm-diameter Al grain (99.99% pure)[23]. A scanning electron micrograph (SEM) of a cantilever with attached Al grain is shown in Fig. 1(b). We drive the cantilever in a phase-locked loop and measure $f_0$ as a function of applied magnetic field $H$ as shown in Fig. 3(a). The red data (positive sweep of $H$) and blue data (negative sweep) have been shifted slightly to correct for hysteresis in the magnet.

Figure 3 (a) shows that below a critical field $H_c$ (indicated on the graph), $f_0 \propto H^2$, while above $H_c$, $f_0$ abruptly drops back to its $H$ = 0 value and ceases to depend on $H$. We interperet this jump as the grain's transition from the superconducting state to the normal state. The quadratic dependence of $f_0$ on $H$ in the superconducting state arises from a combination of the grain's Meissner effect, which induces a magnetic moment $m \propto H$, and the grain's non-spherical shape. This combination leads to a dependence of the grain's energy $E$ upon the angle $\theta$ its principal axis makes with $H$. The $\theta$ dependence of $E$ results from the energy associated with the grain's demagnetizing fields $E_D = \frac{1}{2}\mu_0 m^2 N(\theta)$ where $N(\theta)$ is a shape anisotropy factor[24]. The shift in $f_0$ is proportional to $\partial^2 E/\partial\theta^2$ which in turn is proportional to $H^2$. The hysteresis seen in (a) is due to supercooling of the Al particle[25].

Figure 3(b) shows $H_c$ as a function of $T_b$. Fitting the data using BCS theory[26] (which predicts $H_c(T) \approx H_c(0)(1-(\frac{T_e}{T_c})^2)$) yields $H_c(0)$ = 123 G and $T_c(0)$ = 1.19 K. This value of $T_c$ agrees with the value for bulk aluminum. The measured $H_c(0)$ is slightly greater than the bulk value, which may be due to finite size effects, the grain's nonspherical shape, and the presence of trace impurities[27]. The data and fit in Fig. 3 (b) indicate that $T_e$ follows $T_b$ down to 300 mK for $P_{inc}$ = 25 nW.

The data in Figs. 2 (c) and 3 (b) confirm that the cantilever's undriven motion and the Al grain's $H_c$ serve as thermometers for $T_n$ and $T_e$ respectively. We can use these thermometers to measure the response of $T_n$ and $T_e$ to a localized heat source by measuring $\langle x^2 \rangle$ and $H_c$ as a function of $P_{inc}$. Fig. 4 (a) shows $T_n$ and $T_e$ vs. $P_{inc}$ at $T_b$ = 300 mK. For $P_{inc}$ = 25 nW (the value used for the data in Figs. 2 & 3) $T_n$ and $T_e$ are equal to $T_b$, as discussed above. Fig. 4 (a) shows that higher $P_{inc}$ causes heating of $T_n$ and $T_e$ above $T_b$, presumably due to partial absorption of the laser by the cantilever.

We model the data in Fig. 4(a) by assuming the cantilever has a thermal conductance $\kappa(T)$, a fixed temperature $T_b$ at its base, and a heat source $\dot{Q} = \alpha \tau P_{inc}$ at the location of the laser spot, where $\alpha$ is the cantilever's optical absorption coefficient. At the temperatures of our experiment heat transport in the cantilever is dominated by phonons, so we expect $\kappa(T) = bT^3$, where $b$ is a constant. This gives $\dot{Q} = \frac{w\tau b}{L^*}(T_e^4 - T_b^4)$, where $L^* = 400$ μm is the distance between the laser spot and the cantilever base. In Fig 4 (b) we plot the measured $T_e^4 - T_b^4$ as a function of $P_{inc}$. The solid line is a fit to the expression above.

Assuming a typical value of $b = .0125$ W/cm-K[28] gives $\alpha = 7.5$ cm$^{-1}$. This result for $\alpha$ agrees with direct measurements of optical loss in similarly doped Si at cryogenic temperatures[29]. We note that measurements of optical loss typically compare incident optical power with transmitted and reflected power, and so measure the sum of absorption and diffusive scattering. Our result for $\alpha$ is a direct measurement of absorption.

In conlusion, we have measured both the thermomechanical noise temperature and the sample temperature for a sample-on-cantilever system. Both can remain in thermal contact with a bath for temperatures at least as low as 300 mK. Given the signal-to-noise ratio in Fig. 2 (b), this approach could be used with much smaller samples, including microfabricated devices. We also determined the optical absorption of the cantilever.

We acknowledge useful discussions with Michel Devoret.

REFERENCES;

1. J. A. Sidles, J. L. Garbini, K. J. Bruland, D. Rugar, O. Zuger, S. Hoen, and C. S.Yannoni, Rev. Mod. Phys. **67**, 249 (1995).

2. D. Rugar, R, Budakian, H. J. Mamin, and B. Chui, Nature **430**, 329 (2004).

3. J. G. E. Harris, D. D. Awschalom, R. Knobel, N. Samarth, K. D. Maranowski, A. C. Gossard, Phys. Rev. Lett. **86**, 4644 (2001); J. G. E. Harris, R. Knobel, K. D. Maranowski, A. C. Gossard, N. Samarth, and D. D. Awschalom, Appl. Phys. Lett. **82**, 3532 (2003)

4. Y. Wang, L. Li, M. J. Naughton, G. D. Gu, S. Uchida, and N. P. Ong Phys. Rev. Lett. **95**, 247002 (2005).

5. M. P. Schwarz, D. Grundler, I. Meinel, CH. Heyn, and D. Heitmann, App. Phys. Lett. **76**, 3564 (2000).

6. W. Marshall, C. Simon, R. Penrose, and D. Bouwmeester, Phys. Rev. Lett. **91**, 130401 (2003).

7. Fereira, A., Geirreiro, A. & Vedral, V. Macroscopic thermal entanglement due to radiation pressure. *Phys. Rev. Lett.* **96**, 060407 (2006).

8. Pinard, M., Dantan A., Vitali D., Arcizet, A., Briant, T & Heidman, A. Entangling movable mirrors in a double-cavity system. *Europhys. Lett.* **72**, 747-753 (2005).

9. V. Braginsky and S. P. Vyatchanin, Phys. Lett. A **293**, 228 (2002).

10. H. J. Mamin and D. Rugar, App. Phys. Lett. **79**, 3358 (2001).

11. S. Groblacher, S. Gigan, H. R. Bohm, and A. Zeilinger, arXiv:0705.1149v1 [quant-phys].

12. M. D. LaHaye, O. Buu, B. Camarota, and K. Schwab, Science **304**, 74 (2004).
H. J. Mamin and D. Rugar Appl. Phys. Lett. **79**, 3358 (2001).

13. A. Naik, O. Buu, M. D. LaHaye, A. D. Armour, A. A. Clerk, M. P Blencowe, and K. C. Schwab, Science 443, 193 (2006).

14. C. H. Metzger and K. Karrai, Nature 432, 1002 (2004).

15. S. Gigan, H. R. Böhm, M. Paternostro, F. Blaser, G. Langer, J. B. Hertzberg, K. C. Schwab, D. Bäuerle, M. Aspelmeyer and A. Zeilinger Nature **444**, 67 (2006).

16. D. Kleckner and D. Bouwmeester, Nature **444**, 75 (2006).

17. O. Arcizet, P.-F. Cohadon, T. Briant, M. Pinard and A. Heidmann, Nature **444**, 71 (2006).

18. J. D. Thompson, B. M. Zwickl, A. M. Jayich, Florian Marquardt, S. M. Girvin, J. G. E. Harris, arXiv:0707.1724v2 [quant-ph].

19. K. R. Thurber, L. E. Harrell, and D. D. Smith, J. Appl. Phys. **93**, 4297 (2003).

20. H. J. Mamin, M. Poggio, C. L. Degen, and D. Rugar Nature Nanotech. **2**, 301 (2007).

21. Janis Research, Wilmington MA, USA

22. NanoWorld, Neuchâtel, Switzerland

23. ESPI Metals, Ashland OR, USA

24. E.C. Stoner and E.P. Wohlfarth, Philos. Trans. R. Soc. London, Ser. A 240, 599 (1948).

25. T. E. Faber, Proc. Roy. Soc. A. **231**, 353 (1955).

26. J. Bardeen, L.N. Cooper, and J. R. Schrieffer, Phys. Rev. **108**, 1175 (1957).

27. M. H. Devoret, personal communication

28. M. Asheghi, K. Kurabayashi, R, Kasnavi, and K. E. Goodson, J. Appl. Phys. 91, 5079 (2002).

29. P. E. Schmid, Phys. Rev. B **23**, 5531 (1981).

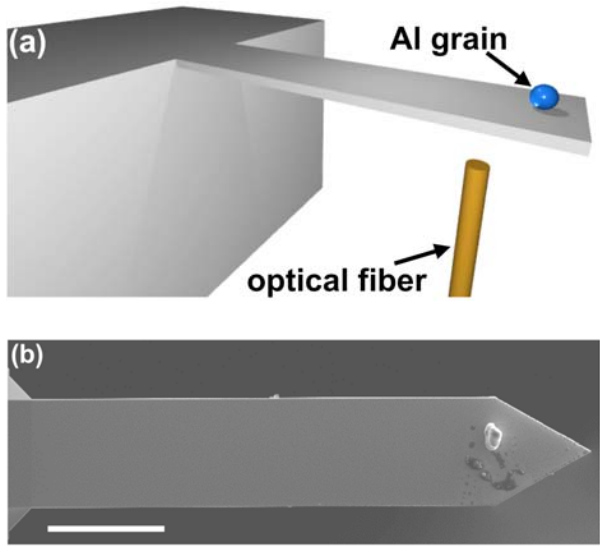

FIG. 1. (a) Experimental schematic. Laser interferometry is used to monitor the deflection of a cantilever (shown here with an Al particle attached). (b) SEM image showing an Al grain epoxied to the end of a Si cantilever. Scale bar is 100 μm.

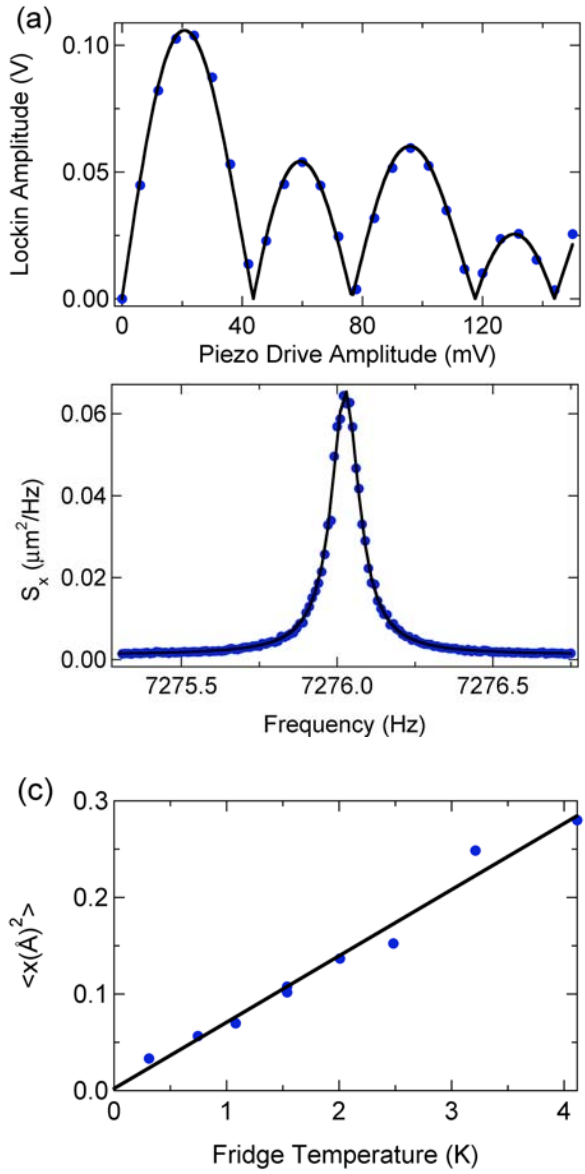

FIG. 2. Cantilever noise thermometry. (a) Calibration of the interferometer signal. The plotted points show the lowest Fourier component of the interferometer signal (as measured by a lock-in amplifier) as a function of the drive amplitude. The fit (solid line) is described in the text. (b) The power spectral density of the cantilever's undriven motion at 4.2 K showing the Brownian motion. (c) Mean square displacement of the cantilever as a function of the refrigerator temperature. The linear fit gives a nearly-zero intercept (solid line), showing the cantilever's undriven motion is thermal.

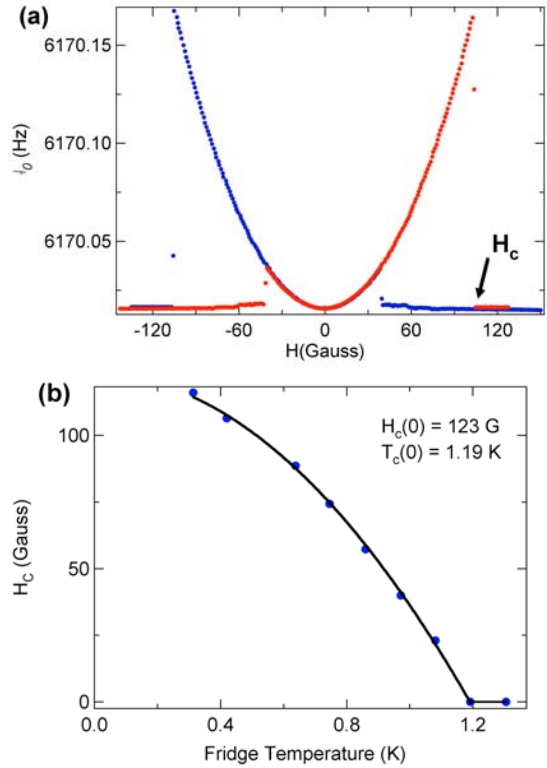

FIG. 3 (color). Thermometry of an aluminum grain attached to the end of a cantilever. (a) Cantilever resonant frequency as a function of applied magnetic field at $T_b = 300$ mK. As the field is swept in the positive direction (red curve), an abrupt jump in $f_0$ occurs at the superconducting critical field $H_c = 110$ Gauss, indicated on the graph. The blue (red) curve is taken during negative (positive) field sweep. (b) Critical field of the Al grain plotted as a function of refrigerator temperature. The blue circles are data and the black curve is a fit to BCS theory.

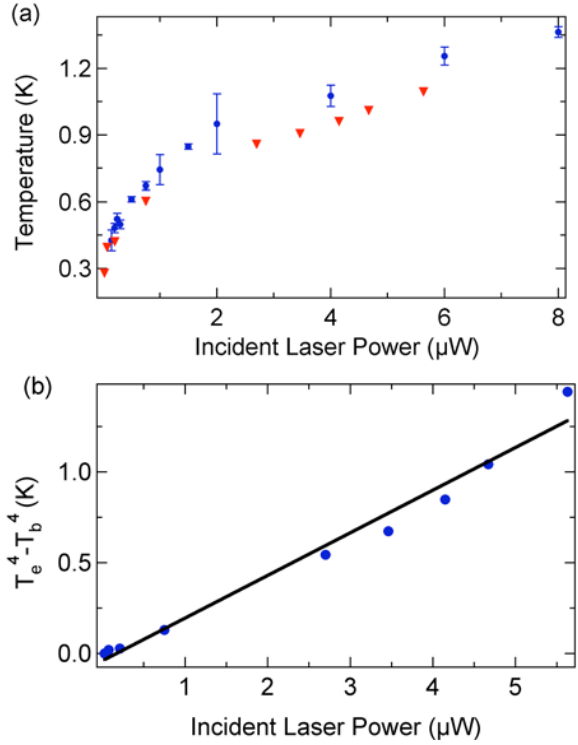

FIG. 4. (a) Electron temperature (red triangles) and noise temperature (blue circles) as a function of laser power incident on the cantilever. The refrigerator temperature is 300 mK at the lowest laser power, and increases slightly for the highest laser powers. (b) The phonon thermal conductivity model described in the text (solid line) is used to fit the temperature difference between the fridge and the aluminum particle (blue dots).